# Magnetic correlations in polycrystalline $Tb_{0.15}Co_{0.85}$

Mathias Bersweiler[1], Philipp Bender[1], Inma Peral[1], Lucas Eichenberger[2], Michel Hehn[2], Vincent Polewczyk[3], Sebastian Mühlbauer[4], and Andreas Michels[1]

[1] Department of Physics and Materials Science, University of Luxembourg, 162A Avenue de la Faïencerie, L-1511 Luxembourg, Grand Duchy of Luxembourg
[2] Institut Jean Lamour (UMR CNRS 7198), Université de Lorraine, 54000 Nancy, France
[3] Istituto Officina dei Materiali (IOM)-CNR, Laboratorio TASC, 34149, Trieste, Italy
[4] Heinz Maier-Leibnitz Zentrum (MLZ), Technische Universität München, D-85748 Garching, Germany

E-mail: mathias.bersweiler@uni.lu, andreas.michels@uni.lu

**Abstract**

We investigated a polycrystalline sample of the ferrimagnetic compound $Tb_{0.15}Co_{0.85}$ by magnetometry and small-angle neutron scattering (SANS). The magnetization curve at 300 K is characteristic for soft ferrimagnets but at 5 K the hysteresis indicates the existence of magnetic domains. The magnetic SANS signal suggests that at 300 K the Tb and Co moments are correlated over large volumes within the micrometer-sized grains with correlation lengths > 100 nm. At 5 K, however, the magnetic SANS analysis reveals a reduced correlation length of around 4.5 nm, which indicates the formation of narrow magnetic domains within the ferrimagnet with one dimension being in the nm range. We attribute the observed changes of the domain structure to the temperature-dependence of the magnetic properties of the Tb sublattice.

## 1. Introduction

Over the last decades ferrimagnetic rare-earth transition-metal alloys (RE-TM) raised a lot of attention, since they are interesting materials for fundamental research [1–4] and promising candidates for technological applications, such as magneto-optical recording media [5], permanent magnets [6], or spintronic devices [7–10]. Previously, all-optical switching has been observed in some selected RE-TM alloys [11–13], rendering them suitable candidates for optically-controlled magnetic data storage devices. More recently, Mangin *et al.* [14] have demonstrated that all-optical helicity-dependent switching can be extended to more complex, multilayered RE-TM systems containing for example HoFeCo, DyCo, or TbCo. Most of the studies are focused on *amorphous* RE-TM alloys, since their fabrication is relatively easy and their magnetic properties can be straightforwardly controlled by changing the concentrations, the nature of RE and TM, or the temperature [15,16]. It is well established that amorphous RE-TM alloys exhibit a noncollinear spin structure, the so-called sperimagnetic structure [17,18], where the magnetic moments are frozen into random orientations. In contrast to the amorphous alloys, in *crystalline* binary intermetallic ferrimagnetic RE-TM alloys, it is assumed that the magnetic moment of RE and TM are antiparallel coupled and form a ferrimagnetic collinear arrangement [19–22].

The goal of the present work is to investigate the structural and magnetic properties of a binary intermetallic Tb-Co ferrimagnetic alloy, one of the most promising candidate system for the next generation of magnetic memories based on all-optical switching [14]. In particular, we study the temperature dependence of the magnetic properties using conventional magnetometry combined with magnetic field-dependent unpolarized small-angle neutron scattering (SANS). Magnetic SANS is a very powerful technique which provides volume-averaged information about variations of the magnetization vector field on a mesoscopic length scale of ~ 1-300 nm [23,24]. This method has previously been applied to study the structures of magnetic nanoparticles [25–32], soft magnetic nanocomposites [33,34], proton domains [35–37], magnetic steels [38–42], or Heusler-type alloys [43–46]. Here, we aim to estimate the temperature dependence of the magnetic correlation length in a *polycrystalline* bulk Tb-Co ferrimagnetic alloy.

## 2. Methods

The polycrystalline $Tb_{0.15}Co_{0.85}$ sample has been prepared by arc melting under a high-purity argon atmosphere starting from stoichiometric quantities of the two high-purity elements (> 99.9% wt. % from Alfa Caesar). The mixture was melted in a water-cooled copper crucible and was not annealed after melting. The sample was then ground manually, compacted to a pellet, and enclosed in a silica tube under purified argon to prevent oxidation. The structural properties were determined by X-ray wide-angle diffraction of the powder using a Bruker D8 DISCOVER diffractometer with a Co-$K\alpha$ radiation source. The magnetic analysis was performed on a pellet using a Physical Property Measurement System (PPMS) from Quantum Design (from 350 K to 5 K in applied magnetic fields up to 4 T). Thermomagnetization $M(T)$ curves and hysteresis loops $M(H)$ were recorded after cooling down the sample under a constant magnetic field of 4 T. The SANS experiments were also performed on a circular pellet, in this case with a diameter of 8 mm and a thickness of $1.2 \pm 0.1$ mm. The neutron experiments were performed at the instrument SANS-1 at the Heinz-Maier-Leibnitz Zentrum (MLZ), Garching, Germany [47]. The measurements were done using an unpolarized incident neutron beam with a mean wavelength of $\lambda = 4.51$ Å and a wavelength broadening of $\Delta\lambda/\lambda = 10\%$ (FWHM). The measurements were conducted at room (300 K) and low temperature (5 K) and within a $q$-range of $0.06$ nm$^{-1}$ $\leq q \leq 3.0$ nm$^{-1}$. A magnetic field $\boldsymbol{H}_0$ was applied perpendicular to the incident neutron beam ($\boldsymbol{H}_0 \perp \boldsymbol{k}_0$). Neutron data were recorded at the maximum field available (4 T) and then in the remanent state (0 T). The neutron-data reduction (correction for background scattering, sample transmission, and detector efficiency) was performed using the GRASP software package [48].

In the neutron data analysis (see below), the *magnetic* SANS cross section is discussed when the total (nuclear + magnetic) SANS cross section at the highest field (near to saturation) is subtracted from the total cross section at a lower field. This procedure assumes that the nuclear SANS cross section is independent of the applied magnetic field. Therefore, it is useful to explicitly display the total and the purely magnetic (difference) SANS cross sections.

When the applied magnetic field is perpendicular to the incident neutron beam ($\boldsymbol{H}_0 \perp \boldsymbol{k}_0$), the elastic total (nuclear + magnetic) unpolarized SANS cross section d$\Sigma$/d$\Omega$ and the purely magnetic SANS cross section d$\Sigma_M$/d$\Omega$ are given as:

$$\frac{d\Sigma}{d\Omega}(\boldsymbol{q}) = \frac{8\pi^3}{V} b_H^2 \Big( b_H^{-2}|\widetilde{N}|^2 + |\widetilde{M}_x|^2 + |\widetilde{M}_y|^2 \cos^2(\theta) \\ + |\widetilde{M}_z|^2 \sin^2(\theta) \\ - \big(\widetilde{M}_y \widetilde{M}_z^* + \widetilde{M}_y^* \widetilde{M}_z\big) \sin(\theta)\cos(\theta)\Big) \quad (1)$$

$$\frac{d\Sigma_M}{d\Omega}(\boldsymbol{q}) = \frac{8\pi^3}{V} b_H^2 \Big(\Delta|\widetilde{M}_x|^2 + \Delta|\widetilde{M}_y|^2 \cos^2(\theta) \\ + \Delta|\widetilde{M}_z|^2 \sin^2(\theta) \\ - \Delta\big(\widetilde{M}_y \widetilde{M}_z^* + \widetilde{M}_y^* \widetilde{M}_z\big) \sin(\theta)\cos(\theta)\Big) \quad (2)$$

where $V$ is the scattering volume, $b_H = 2.91 \times 10^8$ A$^{-1}$m$^{-1}$ relates the atomic magnetic moment to the atomic magnetic scattering length, $\widetilde{N}(\boldsymbol{q})$ and $\widetilde{\boldsymbol{M}}(\boldsymbol{q}) = [\widetilde{M}_x(\boldsymbol{q}), \widetilde{M}_y(\boldsymbol{q}), \widetilde{M}_z(\boldsymbol{q})]$ represent the Fourier transforms of the nuclear scattering length density $N(\boldsymbol{r})$ and of the magnetization vector field $\boldsymbol{M}(\boldsymbol{r})$, respectively, $\theta$ specifies the angle between $\boldsymbol{H}_0$ and $\boldsymbol{q} \cong q\{0, \sin(\theta), \cos(\theta)\}$ in the small-angle approximation, and the asterisks "*" denote the complex conjugated quantities. For small-angle scattering the component of the scattering vector along the incident neutron beam, here $q_x$, is smaller than the other two components, so that only correlations in the plane perpendicular to the incoming neutron beam are probed. The $\Delta$'s in equation (2) represent the difference between the Fourier components at a certain applied field and the highest field of 4 T, which is subtracted in the data analysis. More details about the magnetic SANS technique can be found in Refs. [23,49].

## 3. Results

### 3.1 XRD and SEM

Figure 1 shows the X-ray diffraction results and displays a scanning electron microscopy (SEM) images of the powder. The SEM images show that the primary particles (i.e., grains) are several μm in size. The XRD analysis confirms that $Tb_{0.15}Co_{0.85}$ crystallizes in the hexagonal $CaCu_5$ structure-type with the space group P6/mmm, indicating a pure single phase $TbCo_5$. This is expected from the hypothetical phase diagram of $Tb_xCo_{1-x}$ and for a composition of $x = 0.15$ [50]. Moreover, the XRD pattern exhibits no impurity peaks, which confirms the high-quality synthesis of the Tb-Co alloy by arc melting. The lattice-parameter values $a$ and $c$ were determined by the Le Bail fit method (LBF) implemented in the Fullprof software [51]. The values obtained from the XRD refinement ($a \approx 0.493$ nm and $c \approx 0.401$ nm) are consistent with the values typically obtained in $TbCo_5$ alloy [50,52]. Furthermore, no additional broadening of the diffraction peaks (apart from instrumental broadening) are observed which verifies that the crystallites are at least 100 nm in size.

## 3.2 Magnetometry

Figure 2(a) shows the magnetization curves at 300 K and 5 K. At 300 K, the measured hysteresis loop is similar to that expected for a soft polycrystalline ferrimagnet with randomly distributed anisotropy axis. By cooling down to 5 K, the magnetization curve significantly changes, namely, the magnetization is strongly reduced over the whole field range as compared to 300 K, and the shape of the hysteresis is distinctly different.

The characteristic shape of the hysteresis measured at 5 K (zoom in figure 2(b)) indicates that the reversal becomes dominated by the nucleation (i.e., the jump of $M$ at small reversal fields) and propagation of magnetic domains (i.e., the shearing of the hysteresis at intermediate fields). In fact, the magnetization curve is qualitatively similar to that obtained in synthetic antiferromagnetic magnetic systems whose field reversal behavior has been correlated to the collective propagation of magnetic stripe domains (see figure 3(a) in Ref. [53]).

The temperature dependence of the total magnetization, measured under a cooling-field of 4 T, is displayed in figure 2(c). By decreasing the temperature the total magnetization decreases, as expected by considering the negative exchange coupling between the Co and Tb sublattices (ferrimagnetic) and the temperature dependences of the Co and Tb magnetic moments within the $TbCo_5$ crystal structure. As shown in Ref. [54], in case of Tb-Co alloys having a $TbCo_5$ structure, the magnetization of the Co sublattice remains roughly constant over the temperature range 2-400 K, whereas the magnetization of the Tb sublattice increases significantly with decreasing temperature, which consequently results in a reduction of the total magnetization.

## 3.3 Magnetic Small-angle neutron scattering

Figures 3(a) and (b) display the two dimensional (2D) total (i.e., nuclear + magnetic) SANS cross sections at 300 K and at 5 K, respectively, while figure 4 features the corresponding (over $2\pi$) azimuthally-averaged 1D SANS cross sections. As can be seen, the total 2D SANS cross sections $d\Sigma/d\Omega$ are only weakly field-dependent and isotropic, which suggests the dominance of the isotropic nuclear scattering contribution. According to magnetometry (see figure 2(a)), the sample is nearly magnetically saturated at a field of 4 T for both temperatures. Therefore, assuming a field-independent and isotropic nuclear SANS cross section $d\Sigma_{nuc}/d\Omega$, the 1D sector average of the total SANS cross section parallel to the applied field ($q$ // $H_0$) at 4 T is a good approximation for the nuclear SANS cross section $d\Sigma_{nuc}/d\Omega$ [compare equation (1)]. The in this way estimated 1D $d\Sigma_{nuc}/d\Omega$ [red filled circles in figure 5] exhibit an asymptotic $q^{-4}$ Porod behavior at the smallest momentum transfers. This indicates scattering due to large-scale structures (e.g., grains or pores), which lie outside of the experimentally accessible $q$-range ($> 2\pi/q_{min} \cong 100$ nm). This is expected for the studied sample that consists of μm sized grains (compare the results of the XRD analysis in figure 1).

The 2D magnetic SANS cross section $d\Sigma_M/d\Omega$ in the remanent state at both temperatures [compare Eq. (2)] is determined by subtracting the 4 T data from the measurements at zero field. This data reduction procedure has already been used to extract the purely magnetic SANS cross section of nanoparticle systems [55,56]. The obtained 2D $d\Sigma_M/d\Omega$ are displayed in figures 3(c) and (d) for 300 K and 5 K, respectively. With reference to equation (2) it is reemphasized that the 2D magnetic cross sections $d\Sigma_M/d\Omega$ contain in the sector perpendicular to the field (vertical sector) the difference of $\widetilde{M}_z(q)$ at zero field and at 4 T. On the other hand, the sector average parallel to the field contains the corresponding differences between the transverse magnetization Fourier coefficients $\widetilde{M}_x(q)$ and $\widetilde{M}_y(q)$ at zero field and at 4 T. Thus, analysis of this sector allows to access the transversal magnetic correlation lengths in the remanent state. The 1D magnetic cross sections at 300 K and 5 K, which are obtained by integration along the field direction over an angular range of ± 20°, are displayed in figure 5. At 300 K, the $q$-dependence of $d\Sigma_M/d\Omega$ is similar to that obtained for $d\Sigma_{nuc}/d\Omega \propto q^{-4}$ (at the smallest momentum transfers). This suggests the presence of large magnetic spin-correlation lengths ($l_C > 100$ nm), lying outside of the measured $q$-range. By contrast, at 5 K, a deviation from the $q^{-4}$ dependence can be discerned below 0.2 nm$^{-1}$. This $q$-dependence can be described using a Lorentzian-squared function (blue solid line in figure 5) from which an estimate for the transversal magnetic correlation length of $l_C = 4.5 \pm 0.3$ nm is obtained. As discussed by Hellman *et al.* [57], a Lorentzian-squared term in magnetic SANS data may be attributed to meandering domain walls with $l_C$ being a measure for the domain size. The magnetization data at 5 K (figure 2(a)) together with the estimated nanoscale transversal correlation length are compatible with this result.

## 4. Discussion

We surmise that the formation of narrow magnetic domains observed at 5 K is connected to the temperature dependence of the magnetic anisotropy in $TbCo_5$. The magnetic anisotropy of RE-TM systems is determined by the magnetic anisotropy of both sublattices (here the Tb and Co sublattices). In Ref. [54], it could be shown that the magnetic properties of the Co sublattice barely change within the temperature range 300–5 K. Therefore, it can be assumed that the temperature-dependency of the magnetic properties of $TbCo_5$ is dominated by the one of the Tb sublattice. The magnetic anisotropy of the Tb sublattice depends of the inter-sublattice exchange energy and on the easy-magnetization direction [58], which are both very sensitive regarding temperature. The magnetization of the Tb sublattice significantly increases with decreasing temperature [54], so that it can be assumed that also the magnetic anisotropy constant $K$ increases accordingly [59]. An increase of $K$ favors the formation of narrow domain walls, since the domain-wall width $\delta_\text{w}$ is proportional to $1/\sqrt{K}$ [60]. As reported for $Tb_2Co_{17}$ [61], narrow domain walls are difficult to move and thus may qualitatively explain the observation of narrow domains at low temperature in $TbCo_5$. At room temperature, on the other hand, the magnetic anisotropy of Tb is expected to be weak due to the increased thermal fluctuation of the magnetic moments of the Tb sublattice and, thus, $\delta_\text{w}$ is expected to become larger than at 5 K (the temperature dependence of the magnetic anisotropy in $TbCo_{5.1}$ suggests that $K$ can vanish around 300 K [62]). Therefore, at 300 K, the system may rather favor a correlated single-domain structure within the grains. This feature qualitatively explains the soft ferrimagnetic behavior of the magnetization and the observation of a large spin-correlation length by magnetic SANS, lying outside of the measured $q$-range at room temperature. Further neutron studies, for instance, magnetic-field-dependent polarized SANS and very small-angle neutron scattering (providing access to lower momentum transfers) are required to shed light on the precise nature of the observed correlation lengths: In agreement with previous neutron work [57], we interpreted the origin of the correlation lengths with the domain size, although it has to be considered that the correlation length could also be attributed to the domain walls.

## 5. Conclusion

To summarize, we employed magnetometry and unpolarized SANS to investigate the structural and magnetic properties of *polycrystalline* samples of the ferrimagnetic alloy $Tb_{0.15}Co_{0.85}$. The XRD analysis confirms the high quality of the synthesis with a single phase $TbCo_5$ as expected for this composition. The magnetometry results suggest a reversal of the magnetization by rotation at 300 K, whereas at 5 K the characteristic shape of the hysteresis indicates the nucleation and propagation of magnetic domains. From the unpolarized SANS measurements, the purely magnetic SANS cross sections in the remanent state were determined by subtracting the scattering patterns measured at a large magnetic field of 4 T. The 1D magnetic SANS cross section parallel to the applied field suggests that at 300 K both the Co and Tb moments are correlated over large distances with correlation lengths of at least 100 nm. At 5 K, on the other hand, analysis of the magnetic SANS signal in terms of a Lorentzian-squared scattering function reveals a reduced correlation length of around 4.5 nm. This result in combination with the magnetization curve indicates the formation of domains within the ferrimagnet with one dimension being in the nm range. Finally, we relate our results to the temperature dependence of the magnetic anisotropy of $TbCo_5$, which is dominated by the Tb sublattice for temperatures below 300 K.

## Acknowledgements

The authors acknowledge the Heinz Maier-Leibnitz Zentrum for provision of neutron beamtime. We also thank the C.C. Magnetism and C.C. X-Gamma of the Institut Jean Lamour (Université de Lorraine) for technical support regarding the magnetometry and XRD experiments. P.B. and A.M. acknowledge financial support from the National Research Fund of Luxembourg (CORE SANS4NCC grant).

**Figure Captions**

**Figure 1.** (a) Comparison of the experimental X-ray diffraction pattern of $Tb_{0.15}Co_{0.85}$ (black circles) to the calculated pattern of $TbCo_5$ (red line). For the analysis, the Le Bail fit method (implemented in the Fullprof software) was used, considering the space group P/6mmm. The "*" indicate the diffraction peaks coming from the $K_\beta$ radiation of the Cobalt source. The bottom black solid line represents the difference between the calculated and observed intensities. (b) Secondary electron scanning electron microscopy images of the grain microstructure of our sample. Here the black color corresponds to the carbon tape used for the discharging.

**Figure 2.** (a) Magnetization curves measured in a field range of ± 4 T at 300 K (black solid line) and 5 K (blue solid line). (b) Zoom of the magnetization curve measured at 5 K in a field range of ± 1 T. The onset of nucleation and propagation of the magnetic domains are sketched by the arrows (1) and (2), respectively. (c) Temperature dependence of the total magnetization under a fixed field of 4 T.

**Figure 3.** (a) and (b) Experimental two-dimensional (2D) total (nuclear + magnetic) unpolarized SANS cross sections $d\Sigma/d\Omega$ measured at 300 K and 5 K, respectively. (c) and (d) Purely magnetic 2D SANS cross sections $d\Sigma_M/d\Omega$ measured at 300 K and 5 K, respectively. The purely magnetic 2D SANS cross sections in the remanent state were obtained by subtracting the total scattering at the (near) saturation field of 4 T from the data at H = 0 T. The applied magnetic field $\boldsymbol{H}_0$ is horizontal in the plane of the detector ($\boldsymbol{H}_0 \perp \boldsymbol{k}_0$). Note that the $d\Sigma/d\Omega$ and $d\Sigma_M/d\Omega$ scales are plotted in polar coordinates ($q$ in nm$^{-1}$, $\theta$ in degree, and the intensity in cts/exposure time).

**Figure 4.** (a) and (b) Azimuthally-averaged 1D total SANS cross sections $d\Sigma/d\Omega$ as a function of the momentum transfer $q$ and at selected applied-field values (see insets) (log-log scale) at 300 K and 5 K , respectively. The error bars of $d\Sigma/d\Omega$ are smaller than the data point size.

**Figure 5.** Red filled circles: nuclear 1D SANS cross section $d\Sigma_{nuc}/d\Omega$ as a function of momentum transfer $q$. Colored filled squares: radially-averaged 1D magnetic SANS cross sections $d\Sigma_M/d\Omega$ along the field direction at 300 K (white filled squares) and at 5 K (blue filled squares). Red dashed line: power law $d\Sigma_{nuc}/d\Omega \propto q^{-4}$. Blue solid line: Lorentzian-squared fit of the transverse scattering contribution at 5 K to determine the magnetic transverse correlation length $l_C$. The $d\Sigma_{nuc}/d\Omega$ was determined by ± 10° horizontal sector averages ($\boldsymbol{q}$ // $\boldsymbol{H}_0$) of the total $d\Sigma/d\Omega$ at an applied magnetic field of $\mu_0 H_0$ = 4T and $T$ = 300 K. The radially-averaged 1D magnetic SANS cross section was determined by ± 20° horizontal sector averages ($\boldsymbol{q}$ // $\boldsymbol{H}_0$) of the 2D magnetic SANS cross section at the remanent state taken from figure 3. Note: the magnetic SANS cross section intensities at 300 K and 5 K have been rescaled to the nuclear 2D SANS cross section intensity for better comparison. (log-log scale)

Figure 1

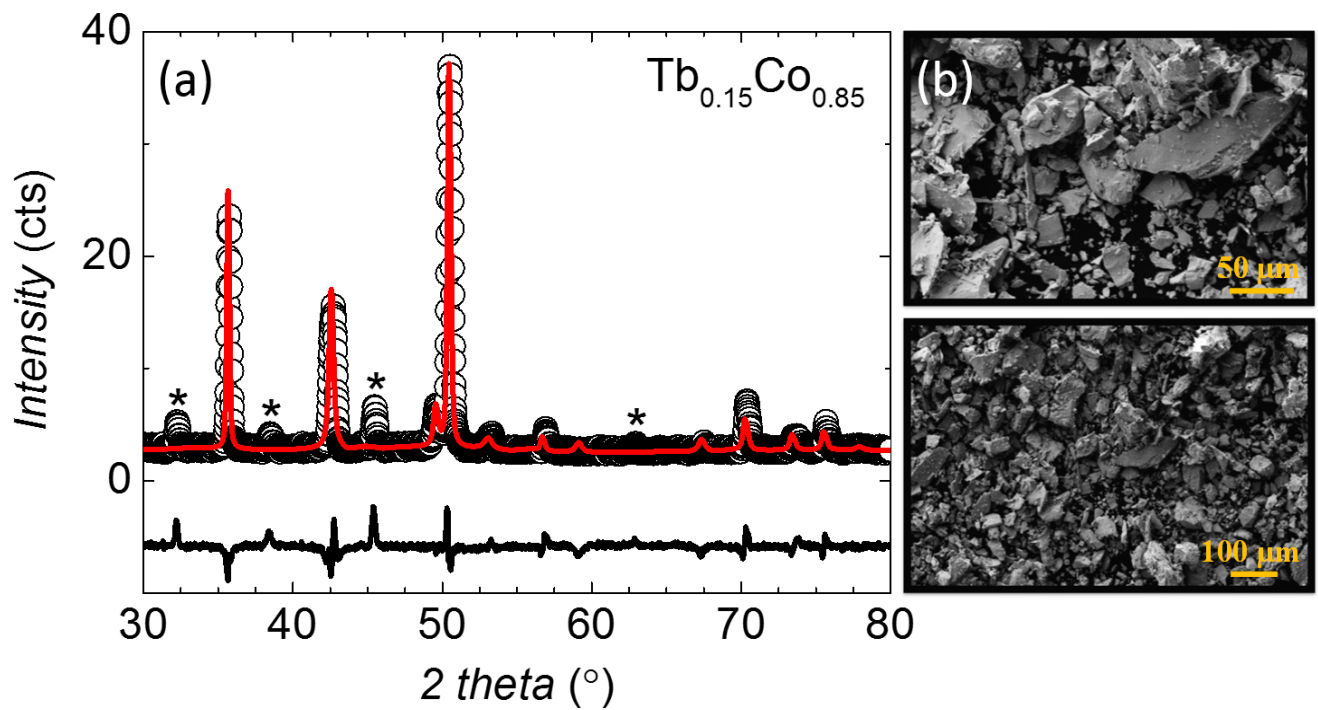

Figure 2

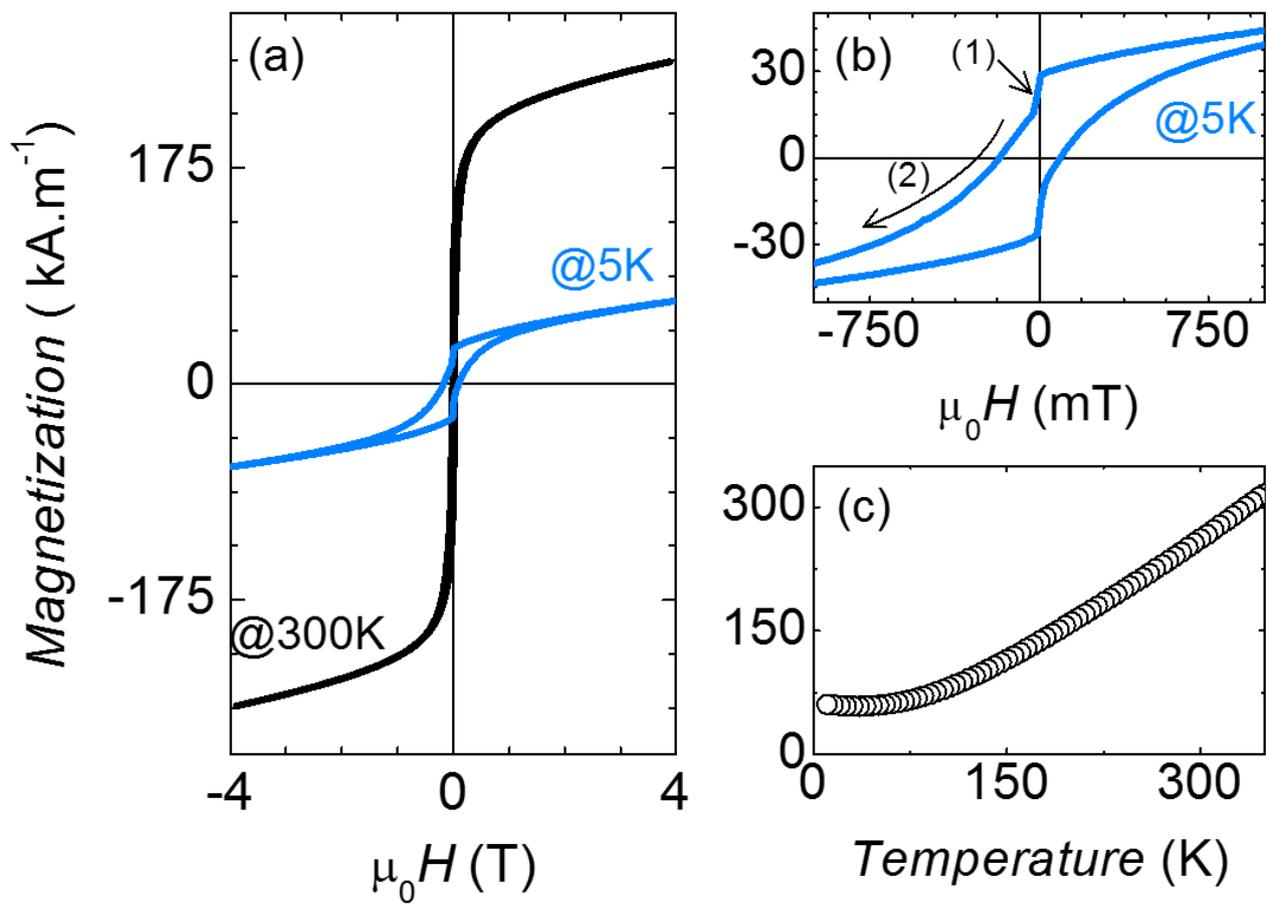

# Figure 3

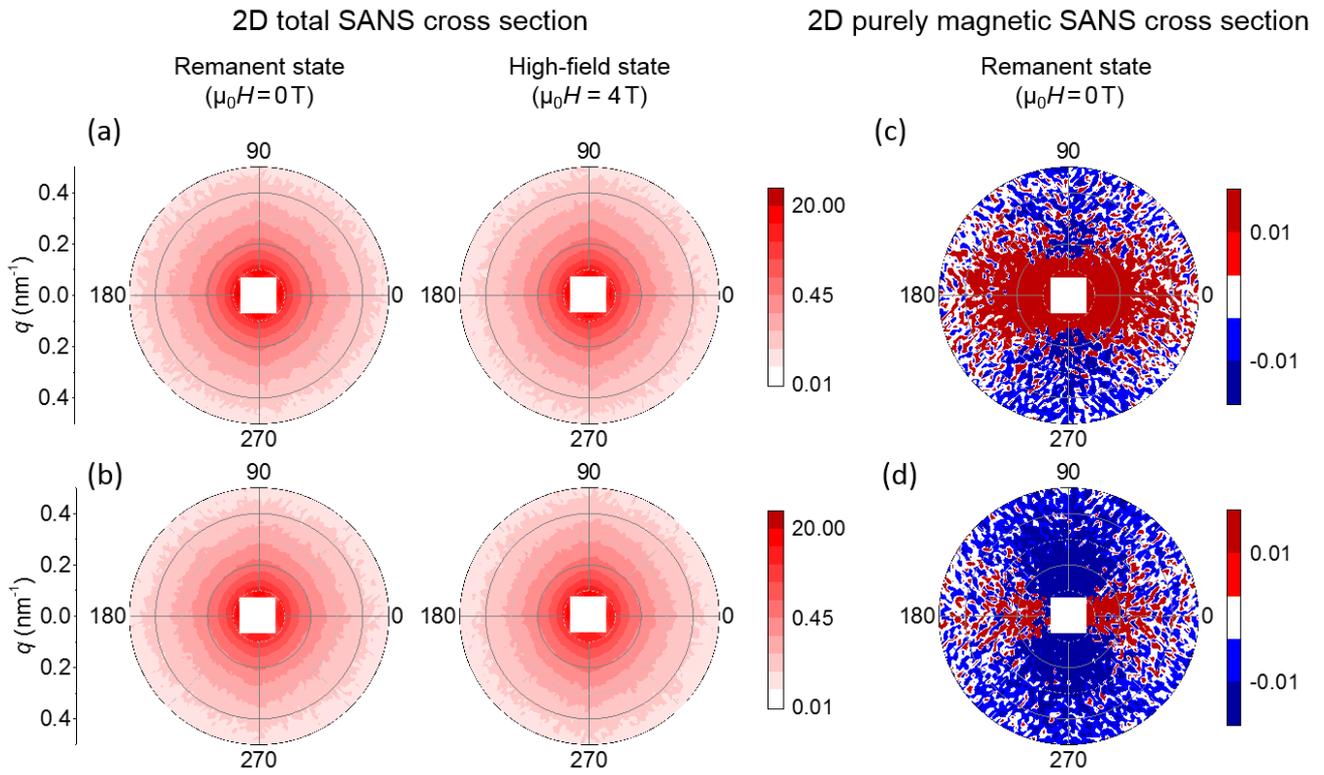

Figure 4

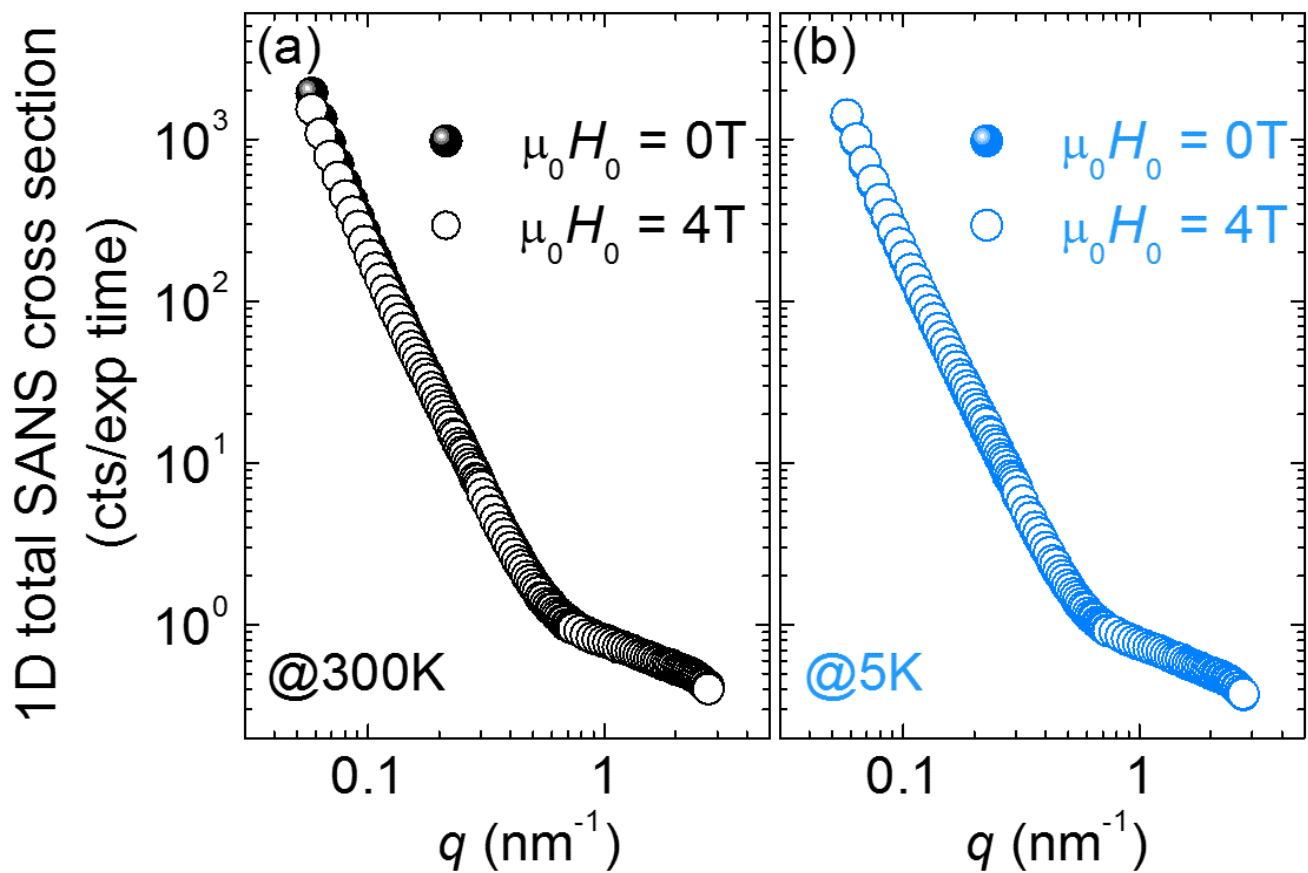

Figure 5

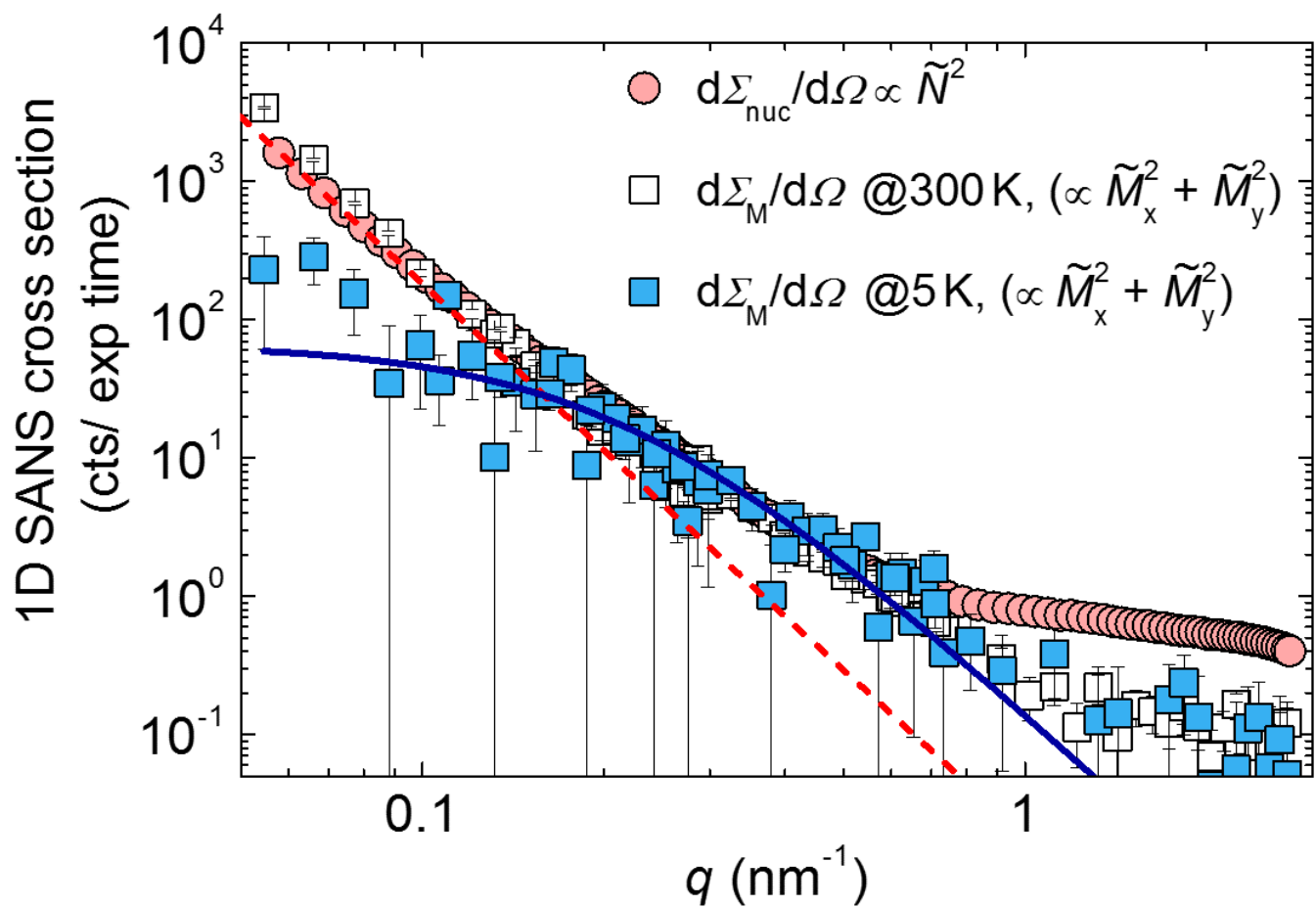